\documentclass{elsart}
\usepackage{amsmath,bm}
\usepackage{graphicx}

\usepackage[normalem]{ulem}  
\usepackage[dvips]{color} 

\begin{document}

\begin{frontmatter}

\title{A direct probe of the in-medium $pn$ scattering cross section}
\author{Gao-Chan Yong,}
\author{Wei Zuo,}
\author{Xun-Chao Zhang}
\address{Institute of Modern Physics, Chinese Academy of
Sciences, Lanzhou 730000, China}

\begin{abstract}

Hard photon production from neutron-proton bremsstrahlung in
intermediate energy heavy-ion reactions is examined as a probe of
the in-medium $pn$ scattering cross section within a transport
model. Uncertainty of photon production probability $pn\rightarrow
pn\gamma$ is cancelled out by using the ratio of hard photon
spectra $R_{^{12}C+^{12}C/p+n}(\gamma)$ from two reactions. The in
medium $pn$ scattering cross section is constrained by using the
ratio of hard photon production cross sections of proton-induced
reactions p+$^{12}$C and p+$^{2}$H. A reduction factor
$\sigma_{pn}^{medium}/\sigma_{pn}^{free}$ of about 0.5 $\sim$ 0.7
around saturation density is obtained by comparing with the
existing experimental data.
\end{abstract}


\end{frontmatter}


The in-medium nucleon-nucleon scattering cross section is a
fundamental physical quantity in nuclear physics and astrophysics
\cite{lat01c,steiner05a,baran05}. Besides many-body theoretical
methods, in fact a lot of literature also reported the studies of
the in-medium nucleon-nucleon scattering cross section based on
transport models. Because so many uncertainties in the transport
model and the complexity in heavy-ion collisions in intermediate
energies, till now the in-medium nucleon-nucleon scattering cross
section is still an open question although the free
nucleon-nucleon scattering cross section is generally considered
as a deterministic quantity. One of the hot topics in today's
nuclear physics is the Equation of State (EoS) of asymmetric
nuclear matter, which is important for understanding many
interesting questions in both nuclear physics and astrophysics
\cite{lat01c,steiner05a,baran05}. Among all the uncertainties of
probing the symmetry energy
\cite{li05,yong11,yong12,yong13,yong14} with heavy-ion collisions,
the nucleon-nucleon scattering cross section is considered to be
one of the most important factors
\cite{li05,yong12,liq2000,ditoro07,zhangy07}. Nowadays almost all
the probes of nucleon-nucleon scattering cross section are
hadronic probes. These probes inevitably suffer from distortions
due to the strong interactions in the final state. Ideally one
expects more clean ways to study nucleon-nucleon scattering cross
section. It is noted that the parity-violating electron scattering
has been proposed to measure more precisely the size of the
neutron-skin in $^{208}$Pb \cite{hor01}. Similarly to electrons,
photons interact with nucleons only electromagnetically. Once
produced they escape almost freely from the nuclear environment of
nuclear reactions. Following the studies of using hard photon
production to probe the symmetry energy \cite{yong08}, in this
Letter, based on the Boltzmann-Uehling-Uhlenbeck (BUU) transport
model, we report our results of using hard photons from
neutron-proton bremsstrahlung in intermediate energy heavy-ion
reactions as a direct probe of the in-medium $pn$ scattering cross
section.

Hard photon production in heavy-ion reactions at beam energies
from about 10 to 200 MeV/nucleon had been extensively studied both
experimentally and theoretically \cite{bertsch88,nif90,cassrp}.
For instance, the TAPS collaboration carried out a series of
comprehensive measurements at various experimental facilities
(GSI, GANIL, KVI) studying in detail the properties of hard
photons in a large variety of nucleus-nucleus systems in the range
of energies spanning $E_{lab}\approx 20-200$ MeV/nucleon
\cite{TAPS}. Theoretically, it was concluded that the
neutron-proton bremsstrahlungs in the early stage of the reaction
are the main source of high energy $\gamma$ rays. Further, it was
demonstrated clearly that the hard photons can be used to probe
the reaction dynamics leading to the formation of dense matter
\cite{bertsch86,ko85,cassing86,bau86,stev86}. Another favorable
factor of using hard photons to probe the $pn$ scattering cross
section is that effects of the nuclear EoS on the hard photon
production was found small \cite{ko87}. One of the major
uncertainties of hard proton studies is the input elementary
$pn\rightarrow pn\gamma$ probability $p_{\gamma}$ which is still
rather model dependent
\cite{nif85,nak86,sch89,gan94,tim06,cassrp}. The recent systematic
measurements of the $pn\rightarrow pn\gamma$ cross sections with
neutron beams up to 700 MeV at Los Alamos and the subsequent
state-of-the-art theoretical investigation may help to improve the
above situation significantly in the near future
\cite{saf07,liy08}.


Since the photon production probability is so small, i.e., only
one photon is produced in roughly a thousand nucleon-nucleon
collisions, a perturbative approach has been used in all dynamical
calculations of photon production in heavy-ion reactions at
intermediate energies \cite{bertsch88,cassrp}. In this approach,
one calculates the photon production as a probability at each
proton-neutron collision and then sum over all such collisions
over the entire history of the reaction. As discussed in detail
earlier in Ref. \cite{cassrp}, the cross section for
neutron-proton bremsstrahlung in the long-wavelength limit
separates into a product of the elastic $pn$ scattering cross
section and a $\gamma$-production probability. The probability is
often taken from the semiclassical hard sphere collision model
\cite{bertsch88,nif90,cassrp}. The single differential probability
reads
\begin{equation}\label{intc}
p^a_{\gamma}\equiv\frac{dN}{d\varepsilon_{\gamma}}
=1.55\times10^{-3}\times\frac{1}{\varepsilon_{\gamma}}(\beta_{i}^{2}+\beta_{f}^{2}).
\end{equation}
Where $\varepsilon_{\gamma}$ is energy of emitting photon,
$\beta_i$ and $\beta_f$ are the initial and final velocities of
the proton in the proton-neutron center of mass frame. We notice
that other expressions derived theoretically involving more
quantum-mechanical effects exist in the literature, see, e.g.,
\cite{nif85,nak86,sch89,gan94,tim06}. For a comparison we thus
also use the prediction of the one boson exchange model by Gan et
al. \cite{gan94}
\begin{equation}\label{QFT}
p^b_{\gamma}\equiv\frac{dN}{d\varepsilon_{\gamma}}=2.1\times10^{-6}\frac{(1-y^{2})^{\alpha}}{y},
\end{equation}
where $y=\varepsilon_{\gamma}/E_{max}$, $
\alpha=0.7319-0.5898\beta_i$, and $E_{max}$ is the energy
available in the center of mass of the colliding proton-neutron
pairs. As noticed already in Ref. \cite{gan94,yong08}, the single
differential probabilities $p^a_{\gamma}$ and $p^b_{\gamma}$ from
the two models give quite similar but quantitatively different
results especially near the kinematic limit where the photon
production with $p^a_{\gamma}$ is significantly higher than that
with $p^b_{\gamma}$. As discussed in the paper of Gan et al.
\cite{gan94}, compared with the result using the semiclassical
expression, the quantum formula $p^b_{\gamma}$ reduces proton
production evidently near the kinematic limit.
\begin{figure}[t]
\begin{center}
\includegraphics[width=0.85\textwidth]{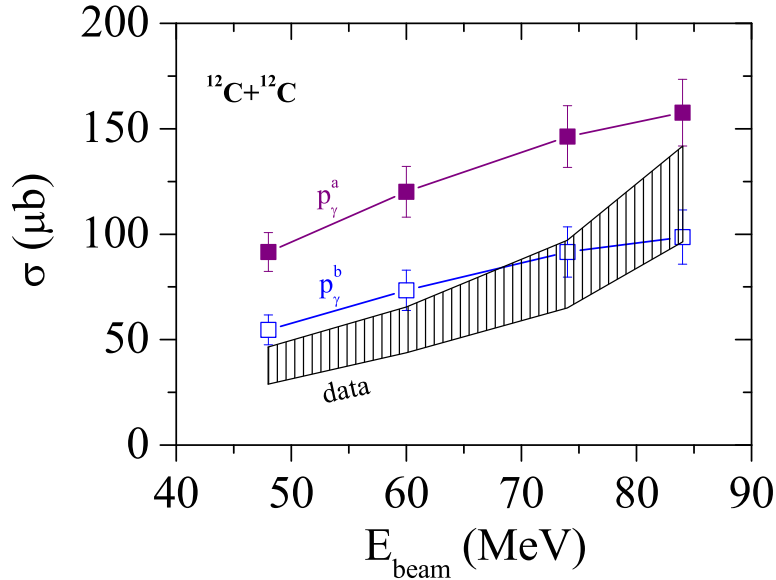}
\end{center}
\caption{(Color online) Beam energy dependence of the inclusive
photon production cross sections in $^{12}$C+$^{12}$C collisions.
The photon energy is 50 MeV $\leq \varepsilon_{\gamma} <$ 100 MeV.
BUU calculations with both $p^a_{\gamma}$ and $p^b_{\gamma}$ vs
experimental data.}\label{vs}
\end{figure}
While we are not aiming at reproducing any data in this
exploratory work, it is necessary to first gauge the model by
comparing with the available data. Shown in Fig.\ \ref{vs} are the
BUU calculations with both $p^a_{\gamma}$ and $p^b_{\gamma}$ and
the experimental data for the inclusive cross sections of hard
photon production in the reaction $^{12}$C+$^{12}$C \cite{yong08}.
It is seen that both calculations are in reasonable agreements
qualitatively with the experimental data, especially at higher
beam energies. The agreement is at about the same level as
previous calculations by others in the literature
\cite{ko85,bau86,gan94}. It is noticed that the uncertainty in the
elementary $pn\rightarrow pn\gamma$ probability leads to an
appreciable effect on the inclusive $\gamma$-production in
heavy-ion reactions. From Fig.\ \ref{vs}, we can clearly see that
the quantum formula $p^b_{\gamma}$ seems more suitable for
describing energetic photon production in intermediate energy
heavy-ion reactions. It was noted that for nuclear bremsstrahlung,
a strong suppression or coherence of the bremsstrahlung cross
section were observed in comparison with predictions of transport
models that include bremsstrahlung on the basis of quasi-free
nucleon-nucleon collisions \cite{goe02,hoe00}. In this study, we
use the BUU transport model \cite{li04a} with an isospin-dependent
in-medium reduced nucleon-nucleon (NN) cross section
\cite{li05,yong12} to treat the in-medium bremsstrahlung
semiclassically \cite{alm95}. The energy and isospin dependent
free-space NN cross sections $\sigma^{free}_{NN}$ are taken from
the experimental data. Another important input to the transport
model is the momentum- and isospin-dependent single nucleon
potential as given in Ref. \cite{das03}. The isoscalar part of the
single nucleon potential was shown to be in good agreement with
that of the variational many-body calculations and the results of
the Brueckner-Hartree-Fock approach including three-body forces
and the isovector potential is consistent with the experimental
Lane potential \cite{yongplb11}.

Since the energetic photons are produced in $pn$ collisions, it is
quite obvious that $pp$ or $nn$ collisions do not influence the
production of photons. We can understand this way: (1) The studied
hard photon is produced via $pn\rightarrow pn\gamma$, so it
reflects $pn$ scatterings directly. The $nn$ or $pp$ scatterings
may affect $pn$ scatterings, and then affect hard photon
production. But they are both secondary effects. (2) The
increased/decreased $pp$ or $nn$ cross sections may also
increase/decrease the collision number of protons and neutrons,
thus may increase/decrease the $pn$ collision number. But, on the
contrary, the larger/small $pp$ or $nn$ cross sections may
reduce/enlarge the collision number of proton with neutron (or
neutron with proton). The secondary effects are thus cancelled out
each other. We in fact checked this deduction by BUU calculations
and find that the hard photon production is really only sensitive
to $pn$ cross section, while not sensitive to $pp$ or $nn$ cross
sections.


\begin{figure}[t]
\begin{center}
\includegraphics[width=0.85\textwidth]{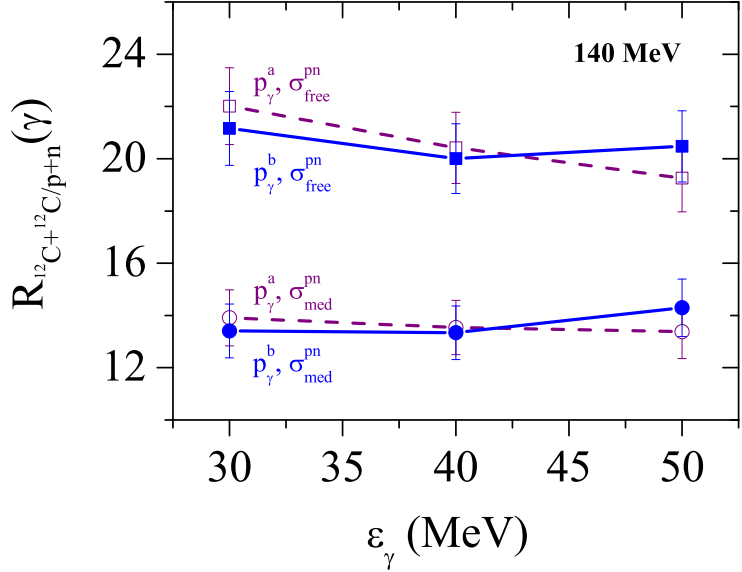}
\end{center}
\caption{(Color online) The ratio of hard photon spectra in the
reactions of $^{12}$C+$^{12}$C and p+n at a beam energy of $140$
MeV/nucleon with free and in-medium $pn$ scattering cross sections
using $p^a_{\gamma}$ and $p^b_{\gamma}$, respectively.}
\label{ratt}
\end{figure}
To reduce uncertainties of the input elementary $pn\rightarrow
pn\gamma$ probability while using hard photon to probe in-medium
$pn$ cross section, we provide Fig. \ref{ratt}, the ratio of hard
photon spectra in the reactions $^{12}$C+$^{12}$C and p+n at a
beam energy of $140$ MeV/nucleon. The spectra ratio
$R_{^{12}C+^{12}C/p+n}(\gamma)$ reads
\begin{equation}
R_{^{12}C+^{12}C/p+n}(\gamma)\equiv\frac{\frac{dN}{d\varepsilon_{\gamma}}(^{12}C+^{12}C)}
{\frac{dN}{d\varepsilon_{\gamma}}(p+n)}. \label{Rphy}
\end{equation}
It is seen that the $R_{^{12}C+^{12}C/p+n}(\gamma)$ is quite
sensitive to the $pn$ scattering cross section while not sensitive
to theoretical formulas used. This ratio reduces the uncertainties
of the theoretical elementary $pn\rightarrow pn\gamma$ probability
maximally. The change of the $pn$ scattering cross section leads
to $R_{^{12}C+^{12}C/p+n}(\gamma)$ a sensitivity of about 60\%.
Like in many experiments searching for minute but interesting
effects, ratio of observables from two reactions can often reduce
not only the systematic errors but also some unwanted effects. At
least theoretically, the uncertainty of the $\gamma$-production
probability gets almost completely cancelled out in the ratio of
photon spectra here. From our BUU calculations, the defined
quantity $R_{^{12}C+^{12}C/p+n}(\gamma)$ does not dependents on
the input $\gamma$-production probability with a high credibility
of 95\%. Also in this definition we assume input
$\gamma$-production probability in matter is the same as that in
vacuum since photons interact with hadrons only
electromagnetically.

The photon production in $^{12}C+^{12}C$ is determined by $pn$
colliding number and the input elementary $pn\rightarrow pn\gamma$
probability, the ratio $R_{^{12}C+^{12}C/p+n}(\gamma)$ thus only
depends on the $pn$ colliding number in the reaction
$^{12}C+^{12}C$ at certain energies. While the $pn$ colliding
number in a reaction depends on the $pn$ cross section in matter.
The spectra ratio $R_{^{12}C+^{12}C/p+n}(\gamma)$ is therefore a
direct probe of the in-medium $pn$ scattering cross section
essentially free of the uncertainties associated with both the
elementary photon production and the $nn$ and $pp$ scattering
cross sections. Compared with other probes of nucleon-nucleon
scattering cross sections, such as nuclear flow and nuclear
stopping, hard photon production directly affects the $pn$
scattering cross section. Practically, besides cosmic-radiation
background, one needs to consider photons from $\pi^{0}$ and
fragment decays in the data analysis \cite{grosse86}.

\begin{figure}[t]
\begin{center}
\includegraphics[width=0.85\textwidth]{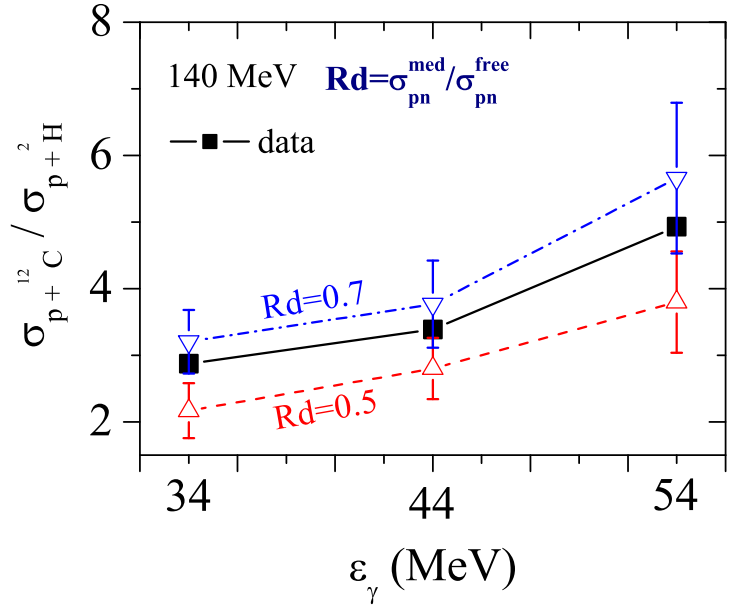}
\end{center}
\caption{(Color online) The ratio of hard photon production cross
sections of p+$^{12}$C and p+$^{2}$H reactions at a beam energy of
$140$ MeV/nucleon with different $pn$ scattering cross sections.
Experimental data are taken from Ref. \cite{bau86}.}
\label{rate12}
\end{figure}
While comparing with the experimental data, we find that there are
few existing data to use. As a rough comparison, we did
simulations of proton induced reactions on $^{12}$C and $^{2}$H
targets at a beam energy of 140 MeV/nucleon as shown in Fig.
\ref{rate12}. Here $\sigma_{p+^{12}C}/\sigma_{p+^{2}H}$ reads
\begin{equation}
\sigma_{p+^{12}C}/\sigma_{p+^{2}H}=\frac{\int_{0}^{b_{max}}\frac{dN}{d\varepsilon_{\gamma}}(p+^{12}C)2\pi
b db}
{\int_{0}^{b_{max}}\frac{dN}{d\varepsilon_{\gamma}}(p+^{2}H)2\pi b
db},
\end{equation}
which in fact is the ratio of $R_{p+^{12}C/p+^{2}H}(\gamma)$ with
different impact parameters. In the calculations we use simple
fermi-momentum as nucleonic momentum in deuteron and $^{12}$C. And
we find that photon production cross section $\sigma_{p+^{2}H}$ is
not sensitive to the distribution of nucleonic momentum in
deuteron \cite{ji89,chen2003,bau86}. We also assumed there is no
medium effect on photon production from p+$^{2}$H. The reference
reaction p+$^{2}$H thus plays roughly the same role as $pn$
collision. We define the reduction factor $Rd=
\sigma_{pn}^{medium}/\sigma_{pn}^{free}$. From Fig. \ref{rate12}
we can see that the experimental data are roughly within the range
of Rd = 0.5 and 0.7 settings of our model. This reduction scale is
somewhat larger than the Brueckner approach calculations
\cite{ligq93,fuchs01,zhf0710}. Experimentally, heavy-ion
collisions with $N\sim Z$ nuclei (to cancel the effect of symmetry
energy) of symmetric system and $p+n$ collision are more suitable
to give constraints on the in-medium $pn$ scattering cross section
by using hard photon production.


In conclusions, we did an exploratory study about effect of the
$pn$ scattering cross section on the production of hard photons
from intermediate energy heavy-ion reactions using a perturbative
approach within the BUU transport model. The ratio of hard photon
spectra $R_{^{12}C+^{12}C/p+n}(\gamma)$ is not only approximately
independent of the uncertainties of $nn$, $pp$ cross sections and
the theoretical elementary $pn\rightarrow pn\gamma$ probability,
but also quite sensitive to the $pn$ scattering cross section.
Compared with other probes of nucleon-nucleon scattering cross
sections, hard photons are completely free of final state strong
interactions, directly reflect the magnitude $pn$ scattering cross
section and are quite sensitive to the $pn$ scattering cross
section. Through comparing with existing experimental data, we
obtain a reduction factor
$\sigma_{pn}^{medium}/\sigma_{pn}^{free}$ of about 0.5 $\sim$ 0.7
around saturation density. Heavy-ion collisions with $N\sim Z$
nuclei of symmetric system and p+n collision are needed to further
constrain the in-medium $pn$ scattering cross section at different
densities and nucleonic momenta.


The author Gao-Chan Yong acknowledges Bao-An Li and Lie-Wen Chen
for the comments on the manuscript. The work is supported by the
National Natural Science Foundation of China (10875151,
10740420550), the Knowledge Innovation Project (KJCX2-EW-N01) of
Chinese Academy of Sciences, the Major State Basic Research
Developing Program of China under No. 2007CB815004, and the
CAS/SAFEA International Partnership Program for Creative Research
Teams (CXTD-J2005-1).

\end{document}